\documentclass[runningheads]{llncs}

\usepackage[T1]{fontenc}
\usepackage{graphicx}
\usepackage{amssymb}
\usepackage{hyperref}

\usepackage{color}

\usepackage{algorithm}
\usepackage{algpseudocode}
\usepackage{amsmath}
\usepackage{pgfplots}
\usepackage{tikz-3dplot}

\pgfplotsset{compat=1.18}

\hypersetup{pdftitle={Vector space embeddings of modal logic formulas: theory and application},
  pdfauthor={Pietro Casavecchia, Alberto Paparella, Leonardo Serrentino}}

\begin{document}

\title{Vector space embeddings of modal logic formulas: theory and application}
\titlerunning{Vector space embeddings of modal logic formulas}

\author{
Pietro Casavecchia\inst{1}\orcidID{0009-0000-9117-234X} \and
Alberto Paparella\inst{1}\orcidID{0009-0007-1653-3660} \and
Leonardo Serrentino\inst{1}\orcidID{0009-0004-9884-0194}
}

\authorrunning{P. Casavecchia et al.}

\institute{
University of Ferrara, Ferrara, Italy \\
\email{\{pietro.casavecchia, alberto.paparella,
leonardo.serrentino\}@unife.it}
}

\maketitle
{\let\thefootnote\relax\footnotetext{Presented at the Student Session of the 36th European Summer School in Logic, Language and Information (ESSLLI 2025), Bochum, 28 July to 8 August 2025. An extended version appeared as \emph{Semantic Embeddings for Modal Logic (Extended Abstract)}, Proceedings of ICTCS 2026, CEUR-WS Vol-4269, pp. 30--35.}}

\begin{abstract}
Reasoning tasks over modal logics are notoriously difficult, comprising problems
which are often semi-decidable.
Hence, it is not rare for applications leveraging such tasks to exploit
heuristic approaches able to give results in an acceptable amount of time, at
the cost of some inaccuracies.
Some approaches may benefit from vector embeddings, allowing for instance to
involve neural networks and machine learning models in the reasoning process.
This work aims at providing a new way to provide such embeddings specifically
for the modal case, aiming for faster reasoning through both mathematical and
learning techniques.

\keywords{Modal logic \and Vector embeddings \and Automated reasoning.}
\end{abstract}

\section{Introduction}

Vector space embeddings of logic formulas aim to represent symbolic logic as
vectors.
This enables, for example, neural networks and machine learning models to
perform reasoning over logic-based
representations~\cite{SMOLENSKY1990159}.
Such embeddings can be given~\cite{sato2017embeddingtarskiansemanticsvector} or
learned from data~\cite{rocktaschel-etal-2014-low,Wang2016LearningFL}.
More recently, embeddings have also been leveraged for
neurosymbolic tasks~\cite{NEURIPS2020_e43739bb,rule,Badreddine_2022}.
However, there seems to be no literature about embeddings of modal logics.
Therefore, a formal framework for vector space embeddings of modal formulas
would be of great value, providing an essential tool for faster reasoning.

This work describes a novel approach to embed modal logic formulas over a vector
space, with the key idea of interpreting a set of representative modal logic
models as dimensions and hence providing a binary encoding of formulas based on
the models that satisfy them. This shares some ideas with
\cite{Benthem2023}, however following different
goals: while the latter aims at describing the semantic meaning of propositions,
this work focuses on finding a set of non-equivalent representative models that
describes the entire universe of all possible Kripke structures.
Furthermore, some examples of both theoretical and applicative interest are
described, solving classical problems in a heuristic way somewhat similar to
\cite{bresolin2013,harmeling2000solving,rossi2000adaptive,venhuizen2021distributionalformalsemantics}.
Finally, we provide an open-source implementation, also describing how one
can get more insightful embeddings.

\section{Vector space embeddings of modal logic formulas}

\begin{figure}[t]
\centering
\tdplotsetmaincoords{75}{100} % Set 3D view (elevation, azimuth)
\begin{minipage}{0.45\textwidth}
\centering
\begin{tikzpicture}[tdplot_main_coords, scale=1]
    % Main axes
    %\draw[->, thick] (0,0,0) -- (3,0,0) node[anchor=east] {$\mathcal{M}_1$};
    %\draw[->, thick] (0,0,0) -- (0,3,0) node[anchor=north] {$\mathcal{M}_2$};
    %\draw[->, thick] (0,0,0) -- (0,0,3) node[anchor=west] {$\mathcal{M}_3$};
    
    \draw[->, thick] (0,0,0) -- (3,0,0) node[anchor=east] {$M_1$};
    \draw[->, thick] (0,0,0) -- (0,3,0) node[anchor=north] {$M_2$};
    \draw[->, thick] (0,0,0) -- (0,0,3) node[anchor=west] {$M_3$};
    
    % Main vectors
    \draw[->, dashed, thick] (0,0,0) -- (2.5,2.5,2.5) node[pos=1, above] {$\varphi_2$};
    \draw[->, dashed, thick] (0,0,0) -- (2.5,0,2.5) node[pos=1, left]  {$\varphi_3,\varphi_4$};
    \draw[->, dashed, thick]  (0,0,0) -- (2.5,2.5,0) node[pos=1, above right] {$\varphi_1$};
    
    \draw (0,0,0) node[left] {$\varphi_5$};
    
    % Dots representing projections
    \filldraw[black] (2.5,0,0) circle (0.7pt) node[above left] {1};
    \filldraw[black] (0,2.5,0) circle (0.7pt) node[above right] {1};
    \filldraw[black] (0,0,2.5) circle (0.7pt) node[left] {1};
    
    % Projections for vector phi1
    \draw[dotted] (2.5,2.5,2.5) -- (0,0,2.5);
    \draw[dotted] (2.5,2.5,2.5) -- (0,2.5,0);
    \draw[dotted] (2.5,2.5,2.5) -- (2.5,0,0);
    
    % Projections for vector phi2
    \draw[dotted] (2.5,2.5,0) -- (2.5,0,0);
    \draw[dotted] (2.5,2.5,0) -- (0,2.5,0);
    
    % Projections for vector phi3
    \draw[dotted] (2.5,0,2.5) -- (2.5,0,0);
    \draw[dotted] (2.5,0,2.5) -- (0,0,2.5);
    
    % Projection of phi1 onto phi2
    \draw[dotted, thick] (2.5,2.5,2.5) -- (2.5,2.5,0) node[midway, left] {$\varphi_1 \models_e \varphi_2$};
\end{tikzpicture}
\end{minipage}
\hspace{0em}
\tdplotsetmaincoords{85}{100} % Set 3D view (elevation, azimuth)
\begin{minipage}{0.45\textwidth}
\centering
% \vspace{1.5em}
\begin{tikzpicture}[tdplot_main_coords, scale=1.2]
    % Reference vector
    \draw[->, thick] (0,0,0) -- (3,3,3) node[pos=1, right] {$\varphi'$};
    
    % Dashed vectors
    \foreach \x/\y/\z in {
        3/2.8/2.3, 2.5/2.5/2.5, 2/2.5/0.5, 1/1.5/2.8, 2.2/0.8/2.5, 
        0.5/2.5/2, 2.5/0/2.5, 2.7/1.2/2, 1.8/2.8/1.6, 3/2.5/0.2,
        0.8/0.8/3, 2.3/1.7/2.1
    } {
        \draw[->, dashed] (0,0,0) -- (\x,\y,\z);
    }
    
    \draw[-, dashed] (3,3,3) -- (3.65,3,3.65) node[pos=0.5, left] {$\tau$};

    % Cone parameters
    \pgfmathsetmacro{\raggio}{0.7}
    \pgfmathsetmacro{\dx}{3}
    \pgfmathsetmacro{\dy}{3}
    \pgfmathsetmacro{\dz}{3}
    \pgfmathsetmacro{\norma}{sqrt(\dx*\dx + \dy*\dy + \dz*\dz)}
    \pgfmathsetmacro{\ux}{\dx/\norma}
    \pgfmathsetmacro{\uy}{\dy/\norma}
    \pgfmathsetmacro{\uz}{\dz/\norma}
    \pgfmathsetmacro{\altezza}{\norma}
    
    \pgfmathsetmacro{\vx}{1/sqrt(2)}
    \pgfmathsetmacro{\vy}{-1/sqrt(2)}
    \pgfmathsetmacro{\vz}{0}
    
    \pgfmathsetmacro{\wx}{ (\uy*\vz - \uz*\vy) }
    \pgfmathsetmacro{\wy}{ (\uz*\vx - \ux*\vz) }
    \pgfmathsetmacro{\wz}{ (\ux*\vy - \uy*\vx) }
    
    \pgfmathsetmacro{\cx}{\altezza*\ux}
    \pgfmathsetmacro{\cy}{\altezza*\uy}
    \pgfmathsetmacro{\cz}{\altezza*\uz}
    
    % Cone representation
    \foreach \t in {0,5,...,355} {
        \pgfmathsetmacro{\xA}{\cx + \raggio*(cos(\t)*\vx + sin(\t)*\wx)}
        \pgfmathsetmacro{\yA}{\cy + \raggio*(cos(\t)*\vy + sin(\t)*\wy)}
        \pgfmathsetmacro{\zA}{\cz + \raggio*(cos(\t)*\vz + sin(\t)*\wz)}
        
        \pgfmathsetmacro{\xB}{\cx + \raggio*(cos(\t+5)*\vx + sin(\t+5)*\wx)}
        \pgfmathsetmacro{\yB}{\cy + \raggio*(cos(\t+5)*\vy + sin(\t+5)*\wy)}
        \pgfmathsetmacro{\zB}{\cz + \raggio*(cos(\t+5)*\vz + sin(\t+5)*\wz)}
        
        \fill[gray!50, opacity=0.5] (0,0,0) -- (\xA,\yA,\zA) -- (\xB,\yB,\zB) -- cycle;
    }
\end{tikzpicture}
\end{minipage}
\caption{On the left: example of a semantic entailment $\varphi_1\models_e
\varphi_2$ in a 3-dimensional vector space; $\varphi_1$ is the projection of
$\varphi_2$ on the (hyper)-plane formed by the models satisfying $\varphi_1$.
On the right: abstract representation of a high-dimensional vector space;
formulas in the gray region are considered similar to $\varphi'$.}
\label{fig:applications}

\end{figure}
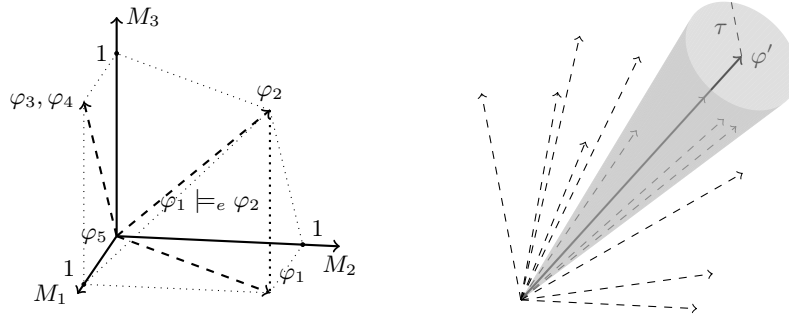

Let $\mathcal{P}=\{p,q,r,\ldots\}$ be a set of propositional letters, $\{\wedge, \vee, \neg\}$ a
set of propositional logic operators, $\{\Box, \Diamond\}$ a set of normal modal
logic operators, $\mathcal{F} = \{\varphi_1, \varphi_2, \dots, \varphi_n\}$ a
set of well-formed formulas, and $\mathcal{M} = \{M_1, M_2, \dots, M_m\}$ a set
of finite Kripke models, where each model $M_i \in \mathcal{M}$ is defined as:
\[ M_i = (W_i, R_i, V_i) \]
where $W_i=\{w_1, ...,w_z\}$ is the set of worlds in $M_i$, $R_i \subseteq W_i
\times W_i$ is an accessibility relation between worlds, and $V_i$ is
a valuation function assigning a truth value $V_i: W_i \times P \to \{0,1\}$ to
each propositional letter in each world.
A vector space embedding for formulas in $\mathcal{F}$ is defined through a
satisfaction matrix $S$ of dimensions $n \times m$, where each element $S_{i,j}
\in \{0,1\}$ is given by:
\begin{equation}
    S_{i,j} =
    \begin{cases} 
        0, & \text{if } M_j, w_1 \not\models_e \varphi_i \\
        1, & \text{if } M_j, w_1 \models_e \varphi_i 
    \end{cases}
    \label{eq:embedding}
\end{equation}

The matrix $S$ summarizes the satisfaction of every formula in $\mathcal{F}$ in
every Kripke structure in $\mathcal{M}$.
Note that $S \subseteq \mathbb{S}$, where $\mathbb{S}$ is the complete matrix
of every possible formula evaluated on every possible Kripke model.
Such an embedding can be interpreted as a vector space comprised by Kripke
models (the columns of $S$) as axes where formulas (the rows of $S$) are
represented by binary vectors, i.e., where each coordinate is either 0 or 1.
To assess how reliably the embedding captures the core logical properties of
each formula, one could leverage a modal reasoner providing a ground-truth
classification by exploring a broad set of formulas and checking their validity,
unsatisfiability, or satisfiability: the embedding should approximate this
classification by mapping the truth values of each formula across $\mathcal{M}$
into a row of $S$, with a row comprising only ones (resp. zeros) suggesting
validity (resp. unsatisfiability), while a row comprising at least a one should
always correspond to satisfiability.

Given a vector space embedding as described in eq.~\ref{eq:embedding}, one can
be interested in a notion of ``similarity'' between formulas: since there seems
to be no shared definition of similarity, this work proposes to consider a
notion of similarity based on the number of models the formulas share.
A figurative example is represented in \autoref{fig:applications}, on the right.
Such a notion can be grasped by common similarity metrics between vectors, e.g.,
the rather popular cosine similarity and its variants. In particular, we employ
a similarity measure between two formulas $\varphi_i$ and
$\varphi_j$ defined as follows:
\begin{equation}
    \text{sim}(\varphi_i, \varphi_j) = 
    \frac{S(i, \cdot)^T S(j, \cdot)}{\min(||S(i, \cdot)||_1,
    ||S(j, \cdot)||_1)}
    \label{eq:similarity}
\end{equation}
where $||S(i, \cdot)||_1$ represents the number of Kripke structures satisfying
formula $\varphi_i$; in fact, $S(i, \cdot)$ represents the $i$-th row of matrix
$S$, while $||S(i, \cdot)||_1$ represents its 1-norm (also called Manhattan
distance).

Given a vector space embedding as described in eq.~\ref{eq:embedding} and a
notion of similarity such as the one in eq.~\ref{eq:similarity}, one could
easily solve, for example, the problem of stating if a formula entails another,
i.e., $\varphi_i\models_e\varphi_j$ where $\varphi_i\models_e\varphi_j\iff\forall
M\in\mathcal{M}: M,w_1\models_e\varphi_i \Rightarrow M,w_1\models_e\varphi_j$.
In fact:
\begin{equation}
\varphi_i\models_e\varphi_j\iff \text{sim}(\varphi_i,\varphi_j)=1
\wedge\min(||S(i,\cdot)||_1,||S(j,\cdot)||_1)=||S(i,\cdot)||_1   
\end{equation}

An example is given in \autoref{fig:applications}, on the left.
One should notice that this is still a heuristic: there could be a model
$\overline{M}\notin \mathcal{M}:\overline{M},w_1\models\varphi_i\wedge
\overline{M},w_1\not\models\varphi_j$.
Hence, the choice of $\mathcal{F}$ and $\mathcal{M}$ is crucial to get reliable results,
and will be discussed in the next section.

Another application of interest could be solving the minimization problem, which
can be defined as follows: given a formula $\varphi_1$, find the formula
$\varphi_2$ which satisfies exactly the same models (i.e., $\varphi
\models_e\varphi_1\wedge\varphi_1\models_e\varphi$) and has the least number of
symbols. One possible way to solve this problem is considering all formulas:
\begin{equation}
\varphi\in\mathcal{F}: \text{sim}(\varphi_i,\varphi_j)=1\wedge||S(i,\cdot)||_1
=||S(j,\cdot)||_1
\end{equation}
and taking the one with the least number of symbols. This could be taken even a
step further if someone is interested in getting a least number of symbols even
at the cost of the formulas disagreeing on some models.
Should this be the case, one could consider a relaxation of the problem, referred
to as quasi-minimization, setting a similarity threshold $\tau$, and
considering all formulas $\varphi\in\mathcal{F}:\text{sim}(\varphi_1, \varphi)
\geq\tau$ and taking the one with the least number of symbols.

\section{Implementation}

\begin{algorithm}[t]
\caption{\textsc{Generate Embedding}}
\begin{algorithmic}[1]
\Require $m$: target dimension, $depth$: maximum formula depth, $W$: maximum number of worlds
\Ensure $S$: embedding matrix, $\mathcal{F}$: list of formulas, $\mathcal{M}$: list of models

\State Initialize $\mathcal{F}$ with 3 randomly generated modal formulas of depth $\leq depth$
\State Initialize $\mathcal{M}$ with 3 Kripke models, each with at most $W$ worlds
\State Construct initial $3 \times 3$ matrix $S$ by evaluating each formula in $\mathcal{F}$ on each model in $\mathcal{M}$

\For{$i = 3$ \textbf{to} $m-1$}

    \State Generate a new modal formula $\varphi_i$ with depth $\leq depth$
    \State Generate a new Kripke model $M_i$ with at most $W$ worlds
    \State Append $\varphi_i$ to $\mathcal{F}$, and $M_i$ to $\mathcal{M}$

    \State Expand $S$ to size $(i+1) \times (i+1)$

    \State Evaluate $\varphi_i$ on all existing models in $\mathcal{M}$ and insert as new row
    \State Evaluate all existing formulas in $\mathcal{F}$ on $M_i$ and insert as new column
    \State Evaluate $\varphi_i$ on $M_i$ and insert at the new diagonal cell

    \State Optimize $S$, $\mathcal{F}$, and $\mathcal{M}$ using \textsc{Iterative Optimization}

\EndFor

\State \Return $S$, $\mathcal{F}$, $\mathcal{M}$
\end{algorithmic}
\label{alg:gen}
\end{algorithm}

\begin{algorithm}[t]
\begin{algorithmic}[1]
\Require $S$: truth matrix $(n \times m)$, $\mathcal{F}$: set of $n$ formulas, $\mathcal{M}$: list of $m$ models, $depth$: formula depth, $W$: maximum number of worlds
\Ensure $S$ has only unique columns; return updated $\mathcal{F}$, $\mathcal{M}$

\While{some columns of $S$ are identical}

    \For{each model index $j$}
        \State Generate a new Kripke model $M_j$ with at most $W$ worlds
        \State Evaluate all formulas in $\mathcal{F}$ on $M_j$ to get a new column $C_j$
        \If{$C_j$ is not already present in $S$}
            \State Replace column $j$ in $S$ with $C_j$
            \State Update model $\mathcal{M}[j]$ with $M_j$
            \State \textbf{break} to re-check uniqueness of $S$
        \EndIf
    \EndFor

    \If{no unique model was found for any column}
        \For{each formula index $i$}
            \State Generate a new formula $F[i]$ of depth $\leq depth$
        \EndFor

        \For{each model index $j$}
            \State Recompute column $j$ of $S$ using the new formulas and model $\mathcal{M}[j]$
        \EndFor
    \EndIf
\EndWhile

\State \Return $S$, $\mathcal{F}$, $\mathcal{M}$
\end{algorithmic}
\caption{\textsc{Iterative Optimization}}
\label{alg:embedding}
\end{algorithm}

The implementation\footnote{The implementation is open-source and is made
available through the following repository:
\href{https://github.com/aclai-lab/ESSLLI2025}{https://github.com/aclai-lab/ESSLLI2025}.}
follows an iterative
approach to construct the satisfaction matrix $S$ identifying a representative
subset of Kripke structures $\mathcal{M}$ aiming to approximate the diversity of
the space of all possible Kripke structures while maintaining computational
feasibility.
The process (Alg.~\ref{alg:gen}) begins with a small set of formulas and Kripke
models, forming an initial satisfaction matrix $S$. 
Then, $S$ is expanded by iteratively introducing new formulas and models
variations (Alg.~\ref{alg:embedding}) within the universe of all Kripke models
while maintaining meaningful embeddings ensuring that no two columns in $S$ are
identical, removing redundancies and enforcing structural differentiation among
the selected Kripke models. 
Whenever a duplication is detected, either new models or formulas are generated to
increase the expressiveness of $S$.
The process continues until a further expansion does not significantly alter the
structural  diversity of the embedding, indicating that $\mathcal{M}$ has reached a
sufficient approximation of the broader Kripke space, capturing key structural
variations necessary for logical reasoning, or when other termination criteria
(e.g., a timeout) are reached.

Once the optimization is complete, the final satisfaction matrix provides a vector
space embedding of modal logic formulas, where each formula's position is
determined by its truth values across the most representative Kripke structures
in $\mathcal{M}$. 
This embedding can then be leveraged for reasoning tasks or integration with
machine learning models, facilitating a more  efficient and scalable approach to
modal logic computations.

\section{Conclusions and future works}

In this work, a possible embedding of modal logic formulas has been proposed
aiming to solve classical problems, such as minimization and entailment, in a
fast way. One should notice how considering only binary encodings is rather
constraining, resulting in much sparsity in the vector space. Hence, a natural
next step would be generalizing this approach to modal fuzzy logics.

\begin{credits}

\subsubsection{\discintname}
The authors have no competing interests to declare that are
relevant to the content of this article.
\end{credits}

\bibliographystyle{splncs04}
\bibliography{refs} % Entries are in the refs.bib file

\end{document}